\documentclass[prl,aps,amssymb,twocolumn,showpacs]{revtex4} 

\usepackage{graphicx}
\begin{document}
\pagebreak
\title{Heart - Shaped Nuclei: 
Condensation of Rotational-Aligned Octupole Phonons   }
\author{S. Frauendorf}
\affiliation{ISP, Forschungszentrum Dresden-Rossendorf, 
 Dresden, Germany}
\affiliation{ Dept. of Physics, University of Notre Dame, Notre Dame, IN 46556}

\date{\today}

\begin{abstract} 
The  strong octupole correlations in the 
mass region $A\approx 226$ are interpreted
as rotation-induced condensation of octupole phonons having their 
angular momentum aligned with the rotational axis.
Discrete phonon energy and  parity conservation generate
oscillations of the energy difference between
the lowest rotational bands with positive and negative parity. 
Anharmonicities tend to synchronize the the rotation of the
 condensate and the quadrupole shape of the nucleus 
forming a rotating heart shape.           
\end{abstract}

\pacs{21.10.Tg, 23.20.Lv, 25.70.Gh, 27.60+j}

\maketitle


\begin{figure}[tb]
\includegraphics[width=7cm]{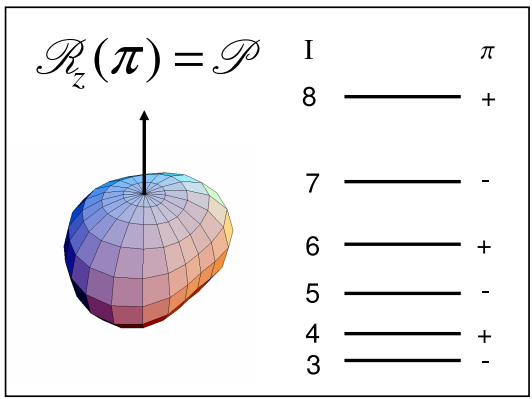}
\caption{Symmetry of a heart-shaped nucleus and the spin-parity
sequence of the rotational ground band (even-even, $\sigma=0$). }
\label{f:symmetry}       
\end{figure}

\begin{figure}[t]
\includegraphics[width=7.5cm]{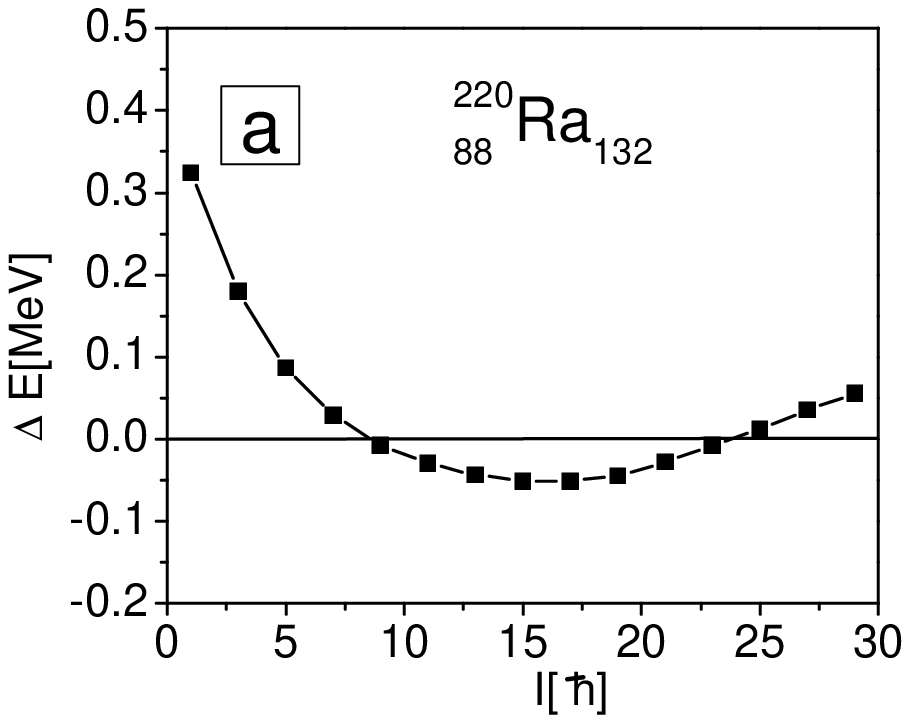}
\includegraphics[width=7.5cm]{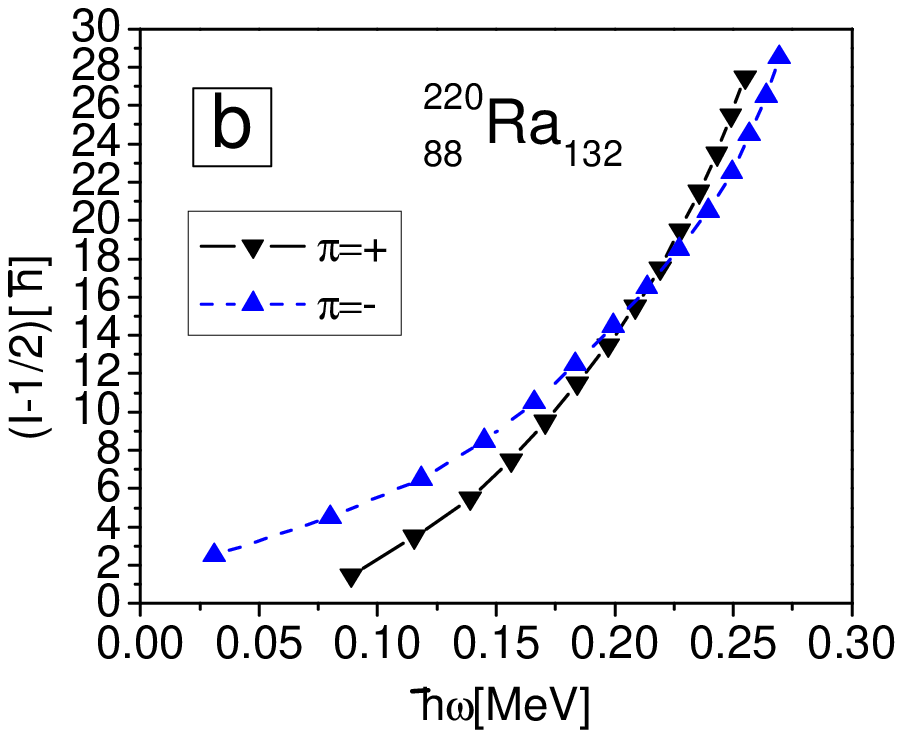}
\caption{a: Energy difference 
$\Delta E(I)=E_-(I)-(E_+(I+1)+E_+(I-1))/2$ 
between the positive and negative parity
yrast sequences in $^{220}$Ra. b: 
Angular momentum $I(\omega)$ as function of the angular frequency
$\hbar\omega(I)=(E(I)-E(I-2))/2$ of the two sequences. 
Data from the ENSDF base.}
\label{f:220Ra}       
\end{figure}

The appearance of rotational bands with alternating parity, as the one
shown in Fig. \ref{f:symmetry}, has been attributed to these nuclei 
having a intrinsic pear shape, which is generated by combining an
axial octupole ($Y_{30}$) with an  axial quadrupole ($Y_{20}$) shape
\cite{ButNaz}. The pear shape has the same symmetry as the more general heart 
shape in Fig. \ref{f:symmetry}.  
The existence of a reflection plane perpendicular to
the rotational axis (z) has the consequence that space inversion $ {\cal P}$
is equivalent to ${\cal R}_z(\pi)$, a rotation by $\pi$ about the
z - axis. The invariance with respect to 
${\cal S}={\cal P \cal R}_z(\pi)$
implies the existence of the simplex quantum number 
${\cal S}|\rangle =e^{-i\pi\sigma}|\rangle $ for the deformed
intrinsic (mean field) state, which leads to the 
alternating spin-parity sequence $\pi=(-1)^{I-\sigma}$ 
of the rotational bands \cite{simplex1,simplex2} (cf. also \cite{RMP}).
The lowest band in  even-even nuclei has $\sigma=0$, as in Fig. 1. 
Strong stretched dipole transitions result from combining 
the octupole distortion  of the shape with the quadrupole
one, which generates a collective charge dipole (cf. eg. \cite{ButNaz}).

For a well developed pear shape one expects that the negative parity states
interleave with the positive parity ones. However in
all alternating bands of even-even nuclei ($\sigma=0$) the negative 
parity sequence is up-shifted relative to the positive parity sequence,
approaching it with increasing spin. This has been interpreted as 
a rapid tunneling mode between the two pear shapes related by $\cal{P}$,
which is progressively suppressed 
with increasing spin and results in merging
of the two sequences \cite{ButNaz,Jolos1,Jolos2}.   
Fig. \ref{f:220Ra}a shows the energy difference  
$\Delta E(I)$ between the $\pi=-$ and $\pi=+$ sequences 
in $^{220}_{88}$Ra$_{132}$.
Obviously, the two sequences do not merge but cross.
Fig. \ref{f:220Ra}b displays the angular momentum as function of the
rotational frequency $\omega$, which is  the slope of $E(I)$. The
$\pi=-$ sequence starts with about $ 3\hbar$ more angular momentum
 than the $\pi=+$ sequence but gains less, such that at high $\omega$ 
the $\pi=+$ sequence has more. 
The conventional interpretation in terms of
pear shape and tunneling does not account for these observations in
a simple way. In this Letter we suggest an alternative interpretation:
Condensation of rotational-aligned octupole phonons.

\begin{figure}[t]
\includegraphics[width=7.4cm]{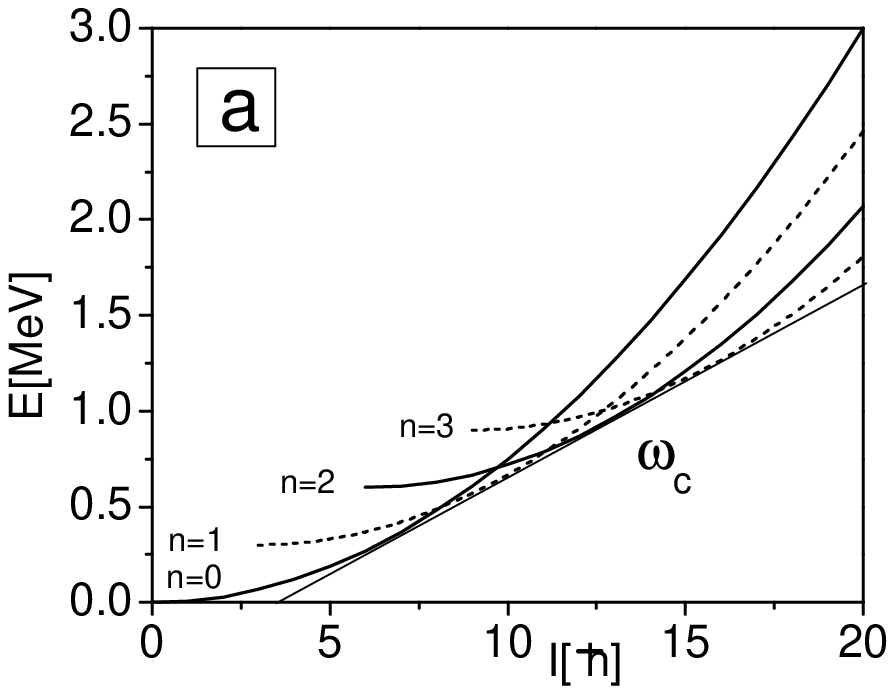}
\includegraphics[width=7.4cm]{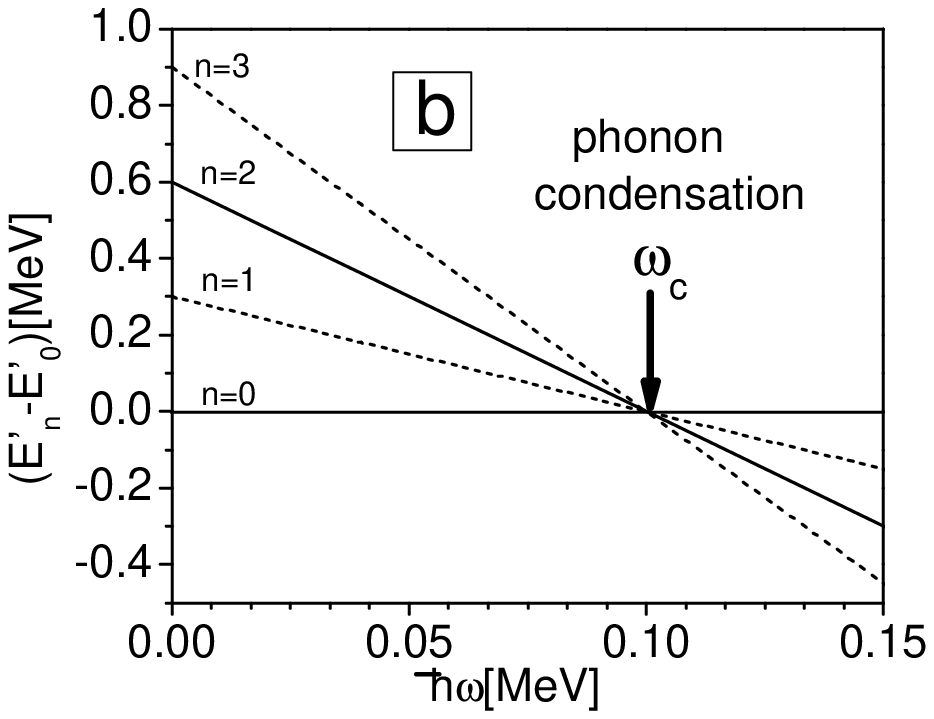}
\includegraphics[width=7.4cm]{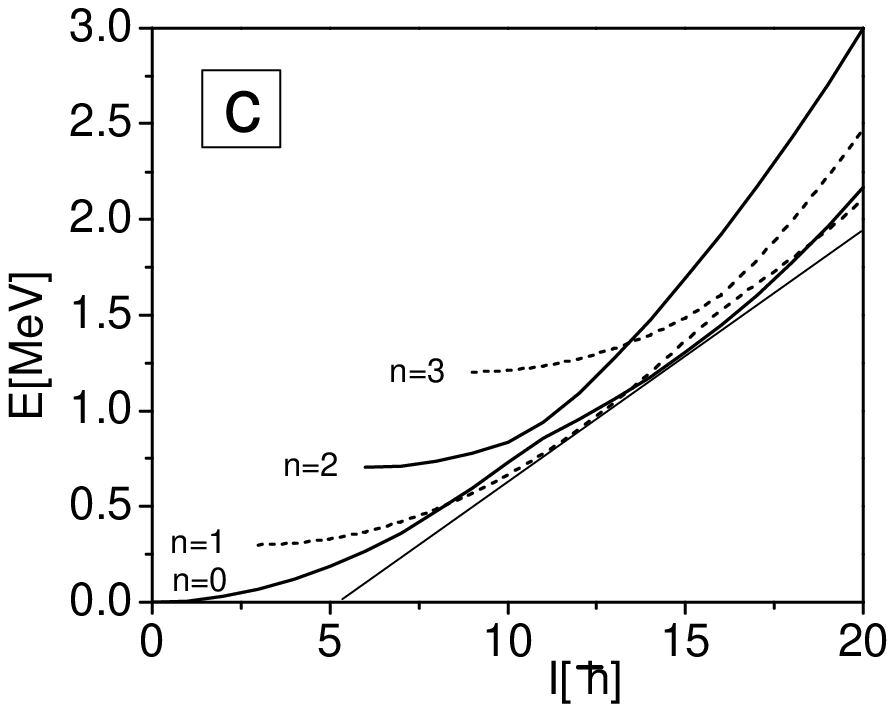}
\caption{a: Energies of aligned octupole multiphonon bands. 
Full lines show $\pi=+$ 
and dashed lines $\pi=-$ states. The straight line
is the common tangent of the curves with the slope $\omega_c$.
b: Energy in the rotating frame (routhian), $E'_n=E_n-\omega I$.
 c: As (a), assuming
an interaction between the phonons and $E_3>3/2E_2>2E_1$.
The slope of the tangent, $\omega$,  increases slowly during condensation. }
\label{f:condens}       
\end{figure}

To present the concept, we assume that the quadrupole deformed
nucleus is a rigid rotor with the moment of inertia 
${\cal J}$, that the octupole vibration is harmonic with frequency 
$\Omega_3$, and that there is no interaction between the octupole
phonons and the quadrupole deformed potential of nucleus. The energy 
of the nucleus in the $n$-boson state 
rotating with the angular velocity $\omega$  is
$E_n=\hbar\Omega_3(n+1/2)+\omega^2{\cal J}/2$, which is
the sum of the boson excitation energy and the rotational energy,
respectively. 
The state with maximal angular momentum for given energy (yrast state)
is generated by aligning the angular momenta of all bosons 
with the axis of rotation. If one boson carries $i\approx 3\hbar$
of angular momentum the total angular momentum of the aligned
$n$-boson state is $I=ni+\omega{\cal J}$ and its energy   
\mbox{$E_n(I)=\hbar\Omega_3(n+1/2)+(I-ni)^2/(2{\cal J})$}, which is shown
in Fig. \ref{f:condens}a. At
$I_n=\hbar\Omega_3{\cal J}/i+i(n+1/2)$  it becomes 
energetically favorable to increase  $I$ by exciting an 
aligned phonon instead of 
further increasing the angular velocity $\omega$. 
Fig. \ref{f:condens}b shows the
energies $E'(\omega)$ of the multiphonon states 
in the frame rotating with the frequency
$\omega$.  These ``routhians'' cross 
at one and the same the critical angular frequency $\omega_c=\Omega_3/i$,
which means there is a boson condensation when the intensive variable 
$\omega$ takes the critical value $\omega_c$.
Fig. \ref{f:condens}a illustrates how the transition shows
up in the relation $E(I)$ between the extensive variables
$E$ and $I$ .
It is spread over many quantal states (the yrast line),
which are distinguished by the discrete variable $I$. 
The energy of these yrast states grows linearly with $I$ {\it on the average},
$\bar E(I)=const+\omega_cI$, 
while the individual energies $E(I)$ fluctuate around it. The critical
frequency $\omega_c$ is the slope of the tangent to the yrast sequence.
 In a macroscopic
system, the fluctuations due to quantization
 become negligible, and one has the  linear
relation $dE/dI=\omega_c$ characteristic for a phase transition, 
which corresponds 
to a vertical section in the function $I(\omega)$ at $\omega_c$. 
In a small system, as the nucleus,
 the values $\omega(I)$ will fluctuate around $\omega_c$   

\begin{figure}[tb]
\includegraphics[width=7.0cm]{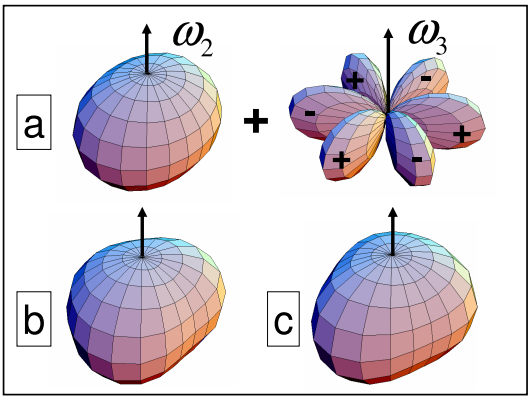}
\caption{An octupole wave traveling over the surface of a 
quadrupole-deformed nucleus. In c, the octupole wave has turned by
90$^\circ$ as compared to b.  }
\label{f:quadoct}       
\end{figure}

The octupole mode in real nuclei shows anharmonicities (cf. e.g.
\cite{BMII}), which let the energy of $n$-phonon states grow faster than
$n\hbar\Omega_3$. In addition, the anharmonicities lead to an 
interaction between the $n$-phonon states,
which causes a repulsion between crossing bands of the {\it same parity},
because due to parity conservation only the states with even $n$ or with
odd $n$ mix.
Fig. \ref{f:condens}c illustrates the condensation  
when these anharmonicities are moderate: The order of the $\pi=+$ and 
$\pi=-$ bands alternates with increasing $I$. Each change of sign 
indicates that one more the phonon has entered the condensate. There is 
no longer a sharp value of $\omega_c$, as discussed for the 
harmonic case. Rather the average angular 
frequency $\omega$ (slope of $E(I)$) increases very slowly
in the condensation region.      

The first steps of the condensation are seen in  $^{220}$Ra. 
Fig. \ref{f:220Ra}a shows the energy difference
$\Delta E=E_--E_+$ between the lowest rotational sequence of each parity.   
The one-phonon band crosses the zero-phonon band  before it feels much of the 
two-phonon band. At the crossing, $\Delta E$
changes sign. When the two-phonon band encounters the zero-phonon one,
the two states mix and exchange character (avoided crossing). The level
repulsion attenuates the growth of $-\Delta E$, which starts decreasing
when the $\pi=+$ band has become predominantly the two-phonon state.
When the two-phonon band crosses  the one-phonon band, $\Delta E$ 
changes sign again. Its growth is attenuated and reversed when the 
avoided crossing between the one- and three-phonon bands is encountered,
the beginning of which is still visible.
Fig. \ref{f:220Ra}b illustrates how the phonon condensation shows
 up in the functions $I(\omega)$. The $\pi=-$
 one-phonon band starts with 
3$\hbar$ more than the $\pi=+$ zero-
phonon band at the same  $\omega$, which is the expected
angular momentum  carried by an aligned octupole phonon. 
The difference decreases, when
the $\pi=+$ two-phonon band, which carries additional  6$\hbar$, 
starts mixing into the zero-phonon band. The (interpolated) $\pi=-$ and $\pi=+$
 bands have equal 
angular momentum at the frequency of maximal mixing
 ($0.5\times 0\hbar+0.5\times 6\hbar=3\hbar$). Near the
 one-two-phonon band crossing  at $I=24$, where the mixing is small,
the angular momentum difference is -3$\hbar$ (at $\hbar\omega=0.26$ MeV).
The other indication for
condensation is the slow growth of $\omega$ above
0.25 MeV, in particular for $\pi=+$. If the phonon spectrum was 
harmonic the two functions
$I_\pm(\omega)$ would be about vertical, 
oscillating around the critical frequency $\omega_c$. It would be
interesting to see if the oscillations of $\Delta E$ and $I(\omega)$,
 which are the 
hallmark of the condensation, continue in experiment
as predicted in Fig. \ref{f:condens}c.

Classically, the $n$-phonon states correspond to an
octupole wave running over the surface of the deformed nucleus
with the angular velocity $\omega_3=\Omega_3/3$ (see Fig. \ref{f:quadoct}). 
The factor 1/3 accounts for the fact 
that the octupole wave reaches an identical position after 
turning 120$^o$. If $\omega=\omega_3$ one has 
a heart shape rotating with $\omega$.
In general, $\omega\not=\omega_3$, which means the
octupole distortion travels with angular velocity $\omega_3-\omega$
relative to the quadrupole-deformed shape (cf. Fig. \ref{f:quadoct}b and c).
At the yrast line, the frequencies $\omega$, which is the slope of 
$E_n(I)$ in Fig. \ref{f:condens}a, 
and $\omega_3=\omega_c$, which is the slope of the tangent,
tend to be equal. However, they cannot completely synchronize because 
the phonon number is discrete. The frequency difference is reflected 
by the fluctuations of
the yrast line above the tangent.
It suppresses the electric dipole 
transitions of the type  $I^+\rightarrow(I-1)^-$. 
The difference $\omega(I^+)-\omega((I-1)^-)=[I-((I-1)-3)]/{\cal J}$ corresponds to a 
decrease  of the angular momentum of the quadrupole rotor by 4$\hbar$.   
The dipole moment is proportional to the product of
the quadrupole moment $Q_2$ and the octupole moment $Q_3$ (c.f. \cite{BMII,ButNaz}).
The transition is suppressed because $Q_2$ can only transfer $\pm 2\hbar$ to the rotor. The
difference $\omega(I^-)-\omega((I-1)^+)=[I-3-(I-1)]/{\cal J}$ corresponds
to an increase of 2$\hbar$, which can be transfered by $Q_2$. The transitions
 $I^-\rightarrow(I-1)^+$ are allowed.

The anharmonicities, which cause
the repulsion between the bands with $n$ and $n+2$ phonons   shown
in Fig. \ref{f:condens}c, attenuate the frequency difference and as 
a consequence the suppression of the $I^+\rightarrow(I-1)^-$ transitions.
For sufficiently strong interaction, 
the $\pi=+$ and $\pi=-$ sequences eventually merge into a smooth sequence
of good simplex $\pi=(-1)^{I-\sigma}$, which  characterizes a static
heart shape. In this case the transitions $I^+\rightarrow(I-1)^-$ and 
$I^-\rightarrow(I-1)^+$ have equal strength, as expected for a rotating static dipole. 

\begin{figure}[tb]
\includegraphics[width=7.5cm]{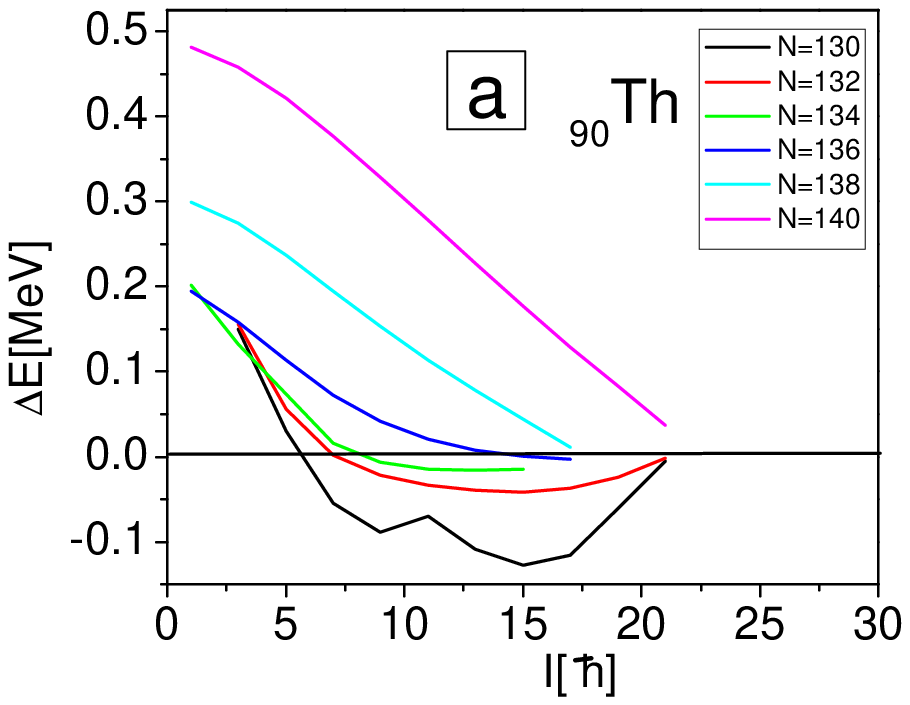}
\includegraphics[width=7.5cm]{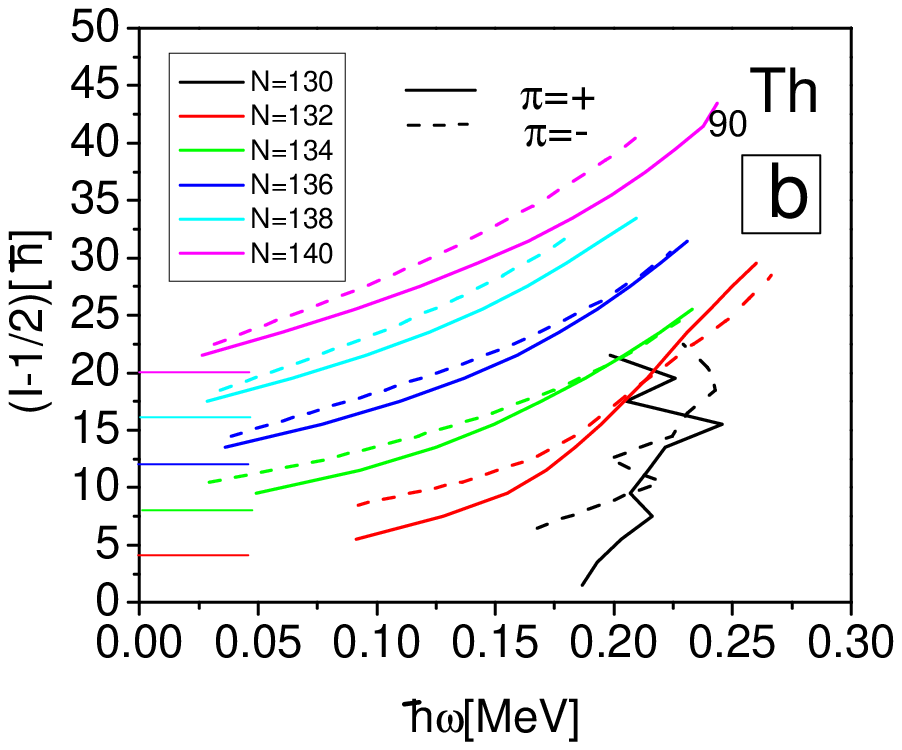}
\caption{As Fig. \ref{f:220Ra} for the Th-isotopes.
Data from the ENSDF base and \cite{reviol2007}. Color online. For more expanded 
versions of b see Fig. 15 of \cite{Cocks99} .}
\label{f:Th}       
\end{figure}

Fig. \ref{f:Th} shows the Th-isotopes. For $N$=130, 132, the $\pi=-$ sequence 
starts with the extra $\approx 3\hbar$ of
the rotational-aligned
octupole phonon. Its distance to the $\pi=+$ sequence, 
$\Delta E(I)$, changes sign
at $I_1=\hbar\Omega_3{\cal J}/i+i/2$.
The mixing and repulsion of the zero- and two-phonon 
bands increases with $N$, which subsequently delays and attenuates the 
 cross-over phenomenon. For $N$=134, the $\pi=-$ sequence 
starts with only $\approx 2\hbar$ of extra angular
momentum, because the $\pi=+$
sequence contains a substantial two-phonon component right from the start, 
which reduces the angular momentum difference with the $\pi=-$ sequence.   
The tendency is continued by $N$=136. The two sequences start with an angular 
momentum difference of only $\approx 1.5\hbar$ and rather merge than cross. 
The progressive synchronization of the angular velocities of the $\pi= +$ 
and - sequences with increasing $N$  signalizes the appearance of a 
static heart shape. The traditional view is that the nucleus attains an 
octupole deformation and there is strong inversion
tunneling, which is suppressed by rotation \cite{ButNaz,Jolos1,Jolos2}. 
The two interpretation do not contradict each other. From the energies
and the $B(E3)$ values of the 3$^-$ states one can estimate
\cite{BMII} that the average elongation of the two-phonon state,
 $\sqrt{<2|\beta_3^2|2>}\sim 0.1$, is comparable with the static deformation
given by mean field calculations around $N=134$ \cite{ButNaz}.  Such a 
deformation is generated by a strong admixture of the two-phonon
and some admixture of the four-phonon state to the zero-phonon state.
However, we argue that the heart shape should be energetically preferred 
over a the pear shape studied so far, which needs to be confirmed by appropriate 
mean field calculations.

The decrease of 
3$^-$ energy and the increase of the 
anharmonicities and phonon coupling with $N$ reflect the Fermi level moving into the 
region where low-$\Omega$ g$_{9/2}$ and j$_{15/2}$ orbitals are close together
(cf.  Fig.5-3 of Ref.\cite{BMII}). The coupling of these orbitals generates
 increasingly soft, anharmonic octupole phonons which easily align with the rotational axis. 
  For $N>136$, the 3$^-$ energy increases strongly with $N$. 
The angular momentum 
difference of $i\approx 3$ near $\Delta E=0$
indicates that the coupling 
between the zero- and two-phonon bands
must be reduced. This can be attributed to
the Fermi level moving from the g$_{9/2}$ to the i$_{11/2}$ orbitals, 
which couple much weaker with the j$_{15/2}$ orbitals via the octupole field 
(cf.  Fig.5-3 of Ref.\cite{BMII}). The larger quadrupole 
deformation in the heavier isotopes increases the quadrupole-octupole coupling,
which favors the $K=0$ octupole phonon. As a consequence, the full alignment 
of the octupole phonon is only reached for $\hbar\omega>$0.15 MeV. 


For $N=130$  
the quadrupole deformation is no longer stable. The 
yrast line is formed by a combined condensation of quadrupole and octupole
phonons. As suggested in \cite{reviol2007},
the rotating condensate forms a heart-shaped  wave running over the nuclear surface.
The condensation is seen in Fig. \ref{f:Th}b as the
roughly vertical $\pi=+$  and $\pi=-$
sequences fluctuating around the critical frequency 
$\hbar\omega_c\approx 0.21$MeV. The $Z=88$ and 86 isotones behave similarly
\cite{Cocks99}.


In summary, the strong octupole correlations of rotational bands 
in the light actinides may be interpreted as the  condensation
 of rotational-aligned octupole phonons.
 During the 
condensation of harmonic 
phonons the energy of  the yrast states increases on the average linearly with 
angular momentum. The discreteness of
the phonon energy combined with parity conservation causes  oscillations of
the lowest positive and negative parity rotational bands
around this classical mean value, which are in anti-phase.
The mismatch of their angular velocities causes a preference 
of electromagnetic dipole transitions  $I^-\rightarrow(I-1)^+$
over $I^+\rightarrow(I-1)^-$.   
The first oscillations of this quantum phase 
transition are clearly seen in the 
$N=132$ isotones. The anharmonicity and interaction
of the phonons increase with $N$, which softens the phase transition and 
 attenuates the oscillations. As a result,  the angular velocities
of the octupole condensate and the quadrupole shape of the nucleus
are progressively synchronized approaching a rotating static heart-shape. 
The $N=136$ isotones are closest to this limit, shows up as 
rotational band of levels with alternating parity that interleave
and equal strength of dipole transitions in both directions. 
For larger $N$ the phonons become again more harmonic.
The strong octupole correlations of rotational bands observed in other mass regions 
can also be interpreted as phonon condensation.

Supported by the DoE Grant DE-FG02-95ER4093.

\end{document}